\documentclass[aps,prd,twocolumn,showpacs,superscriptaddress,%
preprintnumbers,amsmath,amssymb,nofootinbib]{revtex4}

\usepackage{graphicx}% Include figure files
\usepackage{dcolumn}% Align table columns on decimal point
\usepackage{bm}% bold math
\newcommand{\eqn}[1]{&\hspace{-0.3em}#1\hspace{-0.3em}&}
\newcommand{\eV}{\mbox{~eV}}

\newcommand{\GeV}{\mbox{~GeV}}
\newcommand{\TeV}{\mbox{~TeV}}
\begin{document}
%%%%%%%%%%%%%%%%%%%%%%%%%%%%%%%%%%%%%%%%%%%%%%%%%%%%%%%%%%%%%%%%%%%%
%%%%%  Title Page  %%%%%%%%%%%%%%%%%%%%%%%%%%%%%%%%%%%%%%%%%%%%%%%%%
%%%%%%%%%%%%%%%%%%%%%%%%%%%%%%%%%%%%%%%%%%%%%%%%%%%%%%%%%%%%%%%%%%%%
%
%%%%%%%% Preprint #
\preprint{TU-781}
%
%%%%%%%% Title
\title{

Right-Handed Sneutrino as Cold Dark Matter of the Universe

}

%%%%%%%% Authors
\author{Takehiko Asaka}
\affiliation{Institut de Th\'eorie des Ph\'enom\`enes Physiques,
EPFL, CH-1015 Lausanne, Switzerland}
\author{Koji Ishiwata}
\affiliation{Department of Physics, Tohoku University, %
Sendai 980-8578, Japan}
\author{Takeo Moroi}
\affiliation{Department of Physics, Tohoku University, %
Sendai 980-8578, Japan}

%%%%%%%% Date
%\date{\today}
\date{December, 2006}

%%%%%%%%%%%%%%%% Abstract %%%%%%%%%%%%%%%%%%%%%%%%%%%%%%%%%%%%%%%%%%
\begin{abstract}
    We consider the minimal supersymmetric standard model (MSSM)
    extended by introducing three right-handed (s)neutrinos to account
    for neutrino masses in the oscillation experiments.  Assuming that
    the neutrino masses are purely Dirac-type, the lightest
    right-handed sneutrino $\tilde \nu_R$ can be the lightest
    superparticle (LSP), which is a good candidate of cold dark matter
    (CDM) of the universe.  We study the possibility of realizing
    $\tilde \nu_R$-CDM, paying a special attention to the production
    of $\tilde \nu_R$ via decay of the next-to-lightest superparticle
    (NLSP) after its freeze-out time.  It is shown that the late decay
    of the MSSM-LSP (the LSP among superparticles in the MSSM) can
    produce a sufficient amount of $\tilde \nu_R$ to explain the
    observed dark-matter density, and that the $\tilde \nu_R$-CDM
    scenario can be realized in a wide range of parameter space.  We
    also consider the constraint on the decay of MSSM-LSP from the
    big-bang nucleosynthesis (BBN); we found that the case with stau
    being the MSSM-LSP is severely constrained.

\end{abstract}
%%%%%%%%%%%%%%%% Abstract %%%%%%%%%%%%%%%%%%%%%%%%%%%%%%%%%%%%%%%%%%

%%%%%%%% PACS numbers

\maketitle
%%%%%  Title Page  %%%%%%%%%%%%%%%%%%%%%%%%%%%%%%%%%%%%%%%%%%%%%%%%%
%%%%%%%%%%%%%%%%%%%%%%%%%%%%%%%%%%%%%%%%%%%%%%%%%%%%%%%%%%%%%%%%%%%%
%
%%%%%%%%%%%%%%%%%%%%%%%%%%%%%%%%%%%%%%%%%%%%%%%%%%%%%%%%%%%%%%%%%%%%
%%%%%%%%%%%%%%%%%%%%%%%%%%%%%%%%%%%%%%%%%%%%%%%%%%%%%%%%%%%%%%%%%%%%
%%%%% ** Text ** %%%%%%%%%%%%%%%%%%%%%%%%%%%%%%%%%%%%%%%%%%%%%%%%%%%
%%%%%%%%%%%%%%%%%%%%%%%%%%%%%%%%%%%%%%%%%%%%%%%%%%%%%%%%%%%%%%%%%%%%
%%%%%%%%%%%%%%%%%%%%%%%%%%%%%%%%%%%%%%%%%%%%%%%%%%%%%%%%%%%%%%%%%%%%

%%%%%%%%%%%%%%%%%%%%%%%%%%%%%%%%%%%%%%%%%%%%%%%%%%%%%%%%%%%%%%%%%%%%
%%%%%%%%%%%%%%%%%%%%%%%%%%%%%%%%%%%%%%%%%%%%%%%%%%%%%%%%%%%%%%%%%%%%
%%%%%%%%%%%%%%%%%%%%%%%%%%%%%%%%%%%%%%%%%%%%%%%%%%%%%%%%%%%%%%%%%%%%
\section{Introduction}
%%%%%%%%%%%%%%%%%%%%%%%%%%%%%%%%%%%%%%%%%%%%%%%%%%%%%%%%%%%%%%%%%%%%
%%%%%%%%%%%%%%%%%%%%%%%%%%%%%%%%%%%%%%%%%%%%%%%%%%%%%%%%%%%%%%%%%%%%
%%%%%%%%%%%%%%%%%%%%%%%%%%%%%%%%%%%%%%%%%%%%%%%%%%%%%%%%%%%%%%%%%%%%

Supersymmetry is one of the most attractive candidates of physics
beyond the standard model since it may solve various open questions in
particle physics.  It can solve the hierarchy and naturalness problems
and also it can realize gauge-coupling unification.  Supersymmetric
theories have a great advantage in cosmology as well.  This is because
it can provide a viable candidate of cold dark matter (CDM) of the
universe which cannot be understood in the framework of the standard
model of particle physics.  The lightest supersymmetric particle (LSP)
becomes stable with $R$-parity conservation and can be CDM.  Recently,
the relic density of CDM in the present universe has been precisely
determined by the WMAP observation~\cite{Spergel:2006hy}:
\begin{eqnarray}
  \label{eq:omega_dm}
  \Omega_{\mbox{\tiny DM}} h^2 = 0.105^{+0.007}_{-0.013} \,,
\end{eqnarray}
where $h \simeq 0.73$~\cite{Spergel:2006hy} is the present Hubble
constant in units of 100km/sec/Mpc.  Any candidate of CDM must explain
this dark matter density.  Crucial questions left to us are, then,
what is the LSP for CDM and how they are produced accounting for
(\ref{eq:omega_dm}) in the history of the universe.  So far, various
scenarios have been discussed in literatures \cite{Jungman:1995df}.

In Ref.~\cite{Asaka:2005cn}, we have proposed a scenario that the
lightest right-handed sneutrino $\tilde \nu_R$ is the LSP and is CDM
of the universe.\footnote
%%%
{ The possibility of the right-handed sneutrino as CDM has
also been discussed recently in different context.  See
Refs.~\cite{Gopalakrishna:2006kr,McDonald:2006if}.  }
%%%
This is motivated by very small but non-vanishing neutrino masses
strongly suggested by the experimental evidences of neutrino
oscillations (see, {\it e.g.},
Refs.~\cite{SK_S,SK_A,K2K,SNO,KamLAND}). The non-zero neutrino masses
require physics beyond the standard model.  The simplest way to
generate neutrino masses is probably to introduce right-handed
neutrinos (as well as sneutrinos in supersymmetric theories).

It has been widely discussed that right-handed (s)neutrinos are
introduced together with their super-heavy Majorana masses which are
much larger than the electroweak scale $\sim 100$ GeV.  Then the
smallness of neutrino masses is naturally explained by the so-called
seesaw mechanism~\cite{seesaw}.  In this case, the right-handed
sneutrino cannot be the LSP.

On the other hand, neutrinos can obtain very small masses without
invoking the seesaw mechanism.  By introducing right-handed
(s)neutrinos with vanishing Majorana masses, neutrinos become massive
Dirac fermions after the electroweak symmetry breaking.  In this case,
as we will show later, the neutrino Yukawa coupling constants are of
the order of ${\cal O}(10^{-13}-10^{-12})$ or smaller in order to
account for the neutrino mass scales in the oscillation experiments.
It should be noted that such small coupling constants are {\it
natural}\, in the 'tHooft's sense~\cite{thooft}.  This is owing to the
fact that the chiral symmetry in neutrino sector is recovered in the
limit of vanishing neutrino Yukawa coupling constants.

In supersymmetric theories, when neutrino masses are purely
Dirac-type, right-handed sneutrinos receive masses dominantly from the
effects of supersymmetry breaking.  Throughout this paper, we consider
gravity-mediation type models of supersymmetry breaking.  Then, all
the supersymmetry breaking masses are of the order of 0.1--1 TeV.  As
a result, there is a possibility that the lightest right-handed
sneutrino $\tilde \nu_R$ is the LSP.  In Ref.~\cite{Asaka:2005cn}, we
have shown that the LSP $\tilde \nu_R$ is a good candidate of CDM
because it is stable and charge-neutral, and also because its relic
density can become consistent with the dark matter density with
relevant choice of parameters.

In the early universe, $\tilde \nu_R$ is not thermalized since its
interaction is extremely weak.\footnote
%%%
{ If it was thermalized by some unknown interactions in the very
beginnings of the universe, the present relic density of $\tilde
\nu_R$ would overclose the universe and be inconsistent with the
observation.}
%%%
Thus, assuming that the initial abundance of $\tilde \nu_R$ is
zero,\footnote
%%%
{See, however, the discussion in Sec.~\ref{sec:others}.}
%%%
$\tilde \nu_R$ should be produced in some processes after inflation if
$\tilde \nu_R$-CDM is realized.  One possible production is via decays
of superparticles~\cite{Asaka:2005cn}.  Superparticles in the minimal
supersymmetric standard model (MSSM) are in {\it chemical
equilibrium}\ when the temperature is high enough, so they are quite
abundant in the early universe.  Decays of those MSSM superparticles
can produce $\tilde \nu_R$.  The decay processes may occur when the
MSSM superparticles are in the chemical equilibrium and when they
freeze-out from the thermal bath.  Importantly, the decay processes
after the freeze-out may be important for the production of $\tilde
\nu_R$.  This happens for the decay of the lightest superparticle in
the MSSM sector, which we call the MSSM-LSP.  This is because the
neutrino Yukawa coupling constants are so small that the lifetime of
the NLSP becomes rather long.  Such late decay of the NLSP into
$\tilde \nu_R$ can be one important source of $\tilde{\nu}_R$-CDM.

In this paper, we consider the scenario with $\tilde \nu_R$-LSP and
investigate its production in the early universe.  In
Ref.~\cite{Asaka:2005cn}, we have mainly analyzed the production by
decays of superparticles which are in chemical equilibrium.  Here, we
study another production process in detail, that is, decay of the MSSM-LSP
after its freeze-out time.  We estimate the relic density of $\tilde
\nu_R$ from such a decay adopting the minimal supergravity
model~\cite{mSUGRA}, and also discuss the implication of the $\tilde
\nu_R$-CDM scenario.  We also reconsider the decay of MSSM
superparticles in chemical equilibrium.

The outline of this article is the following.  We start by reviewing
the framework of our analysis in Sec.~\ref{sec:framework}, where we
present properties of neutrinos and sneutrinos under consideration.
In Sec.~\ref{sec:production} various production processes of $\tilde
\nu_R$ are discussed.  We first study the production of $\tilde \nu_R$
by decays of the NLSP after the freeze-out time.  The constraint on
the NLSP decay coming from the big-bang nucleosynthesis (BBN) is also
discussed.  We then turn to reconsider the production by decays of
superparticle in chemical equilibrium to complete the discussion in
Ref.~~\cite{Asaka:2005cn}.  In Sec.~\ref{sec:production},
other possible mechanisms of the production of $\tilde \nu_R$
are also mentioned.  Finally, we conclude in
Sec.~\ref{sec:conclusions}.
%%%%%%%%%%%%%%%%%%%%%%%%%%%%%%%%%%%%%%%%%%%%%%%%%%%%%%%%%%%%%%%%%%%%
%%%%%%%%%%%%%%%%%%%%%%%%%%%%%%%%%%%%%%%%%%%%%%%%%%%%%%%%%%%%%%%%%%%%
%%%%%%%%%%%%%%%%%%%%%%%%%%%%%%%%%%%%%%%%%%%%%%%%%%%%%%%%%%%%%%%%%%%%
\section{Framework}
\label{sec:framework}
%%%%%%%%%%%%%%%%%%%%%%%%%%%%%%%%%%%%%%%%%%%%%%%%%%%%%%%%%%%%%%%%%%%%
%%%%%%%%%%%%%%%%%%%%%%%%%%%%%%%%%%%%%%%%%%%%%%%%%%%%%%%%%%%%%%%%%%%%
%%%%%%%%%%%%%%%%%%%%%%%%%%%%%%%%%%%%%%%%%%%%%%%%%%%%%%%%%%%%%%%%%%%%

First of all, let us explain the framework of our analysis.  We
consider the MSSM with three generations of right-handed (s)neutrinos,
where neutrino masses are assumed to be {\it purely Dirac-type}.  The
superpotential is then given by
\begin{eqnarray}
  \label{eq:SP}
  W 
  = y_\nu \, \hat H_u \cdot \hat L \, \hat \nu_R^c 
  - y_e  \, \hat H_d \cdot \hat L \, \hat \ell_R^c
  + \mu_H \, \hat H_d \cdot \hat H_u 
  \,,
\end{eqnarray}
where we have omitted terms with quark superfields since they are
irrelevant for our discussion.  Here $\hat H_u = (\hat H_u^+ , \hat
H_u^0)$ and $\hat H_d = (\hat H_d^0 , \hat H_d^-)$ are the Higgs
superfields coupled to up- and down-type quarks, and $\hat L = ( \hat
\nu_L , \hat \ell_L^-)$ are the left-handed lepton superfields.  The
superfields of right-handed neutrinos are denoted as $\hat \nu_R$.
(In this article, ``hat'' is for superfields while ``tilde'' is for
superparticles with odd $R$-parity.)  $\mu_H$ is the
supersymmetry-invariant Higgs mass.  Here and hereafter, the
generation indices are implicit for simplicity.

In this framework, neutrinos obtain masses as
\begin{eqnarray}
  m_\nu = y_\nu \, \langle H_u^0 \rangle
  = y_\nu \, v \, \sin \beta \,,
\end{eqnarray}
where $v \simeq 174$ GeV is the vacuum expectation value (VEV) of
the standard-model-like Higgs boson, and $\tan \beta = \langle H_u^0
\rangle / \langle H_d^0 \rangle$.  Numerically, we obtain
\begin{eqnarray}
  \label{eq:ynuatm}
    y_\nu \sin\beta \simeq 3.0 \times 10^{-13} \times
    \left( \frac{m_{\nu}^2}{2.8 \times 10^{-3}\ {\rm eV^2}} 
    \right)^{1/2} \,.
\end{eqnarray}
Importantly, neutrino-oscillation experiments have provided only the
mass-squared differences of neutrinos~\cite{K2K,KamLAND}:
%%%
\begin{eqnarray}
  \left[ \Delta m_\nu^2 \right]_{\rm atom}
  \eqn{\simeq} 2.8 \times 10^{-3} \, \mbox{eV}^2 \,,
  \label{m_nu(atom)}
  \\
  \left[ \Delta m_\nu^2 \right]_{\rm solar}
  \eqn{\simeq} 7.9 \times 10^{-5} \, \mbox{eV}^2 \,,
\end{eqnarray}
and the absolute scales of neutrino masses have not been determined.
There are then three possible mass spectra: (i) the normal hierarchy
case, (ii) the inverted hierarchy case, and (iii) the degenerate case.
On the other hand, the cosmological observations place the upper bound
on $m_\nu$.  Here, we adopt the upper bound derived from the WMAP
three-year data: $\sum m_\nu < 2.0$ eV (95\% CL)
\cite{Fukugita:2006rm}, where summation is over all the three neutrino
flavors.\footnote
%%%
{ The inclusion of the data other than the cosmic microwave
background radiation (CMBR) observation ({\it e.g.}, the large scale
structure and the matter power spectrum inferred from the Ly-$\alpha$)
makes the mass limit of neutrinos more stringent (see, for example,
Ref.~\cite{Spergel:2006hy}).  We take here a conservative approach and
use the bound only from the CMBR data.  }
%%%
Then, we obtain
\begin{eqnarray}
    m_\nu < 0.67 \ {\rm eV}.
    \label{upperbound_mnu}
\end{eqnarray}

In the case with hierarchical neutrino masses (i) and (ii), the
heaviest neutrino mass is well approximated by $\sqrt{\left[ \Delta
    m_\nu^2 \right]_{\mbox{\tiny atom}}}$ and the corresponding
neutrino Yukawa coupling constant, which is the largest one, is about
$y_\nu\sin\beta\simeq 3.0\times 10^{-13}$.  On the other hand, in the
degenerate case (iii), the coupling constants become larger; if the
neutrino masses take their largest possible value of $\simeq 0.67\ 
{\rm eV}$, Yukawa coupling constant is as large as $y_\nu \sin \beta
\simeq 3.8 \times 10^{-12}$.  We can see that the neutrino Yukawa
coupling constants are very small in any case, which are much smaller
than those of other quarks and leptons.

Next, we turn to discuss sneutrinos.  For this purpose, let us
introduce the soft supersymmetry breaking terms with sneutrinos as
\begin{eqnarray}
  {}- {\cal L}_{\rm soft}
  \eqn{\supset}
    \tilde M_{L}^2 \, | \tilde L |^2
  + \tilde M_{\nu_R}^2 \, | \tilde \nu_R |^2
  \nonumber \\
  \eqn{}+ \left(
    \tilde A_\nu \, H_u \cdot \tilde L \, \tilde \nu_R^c
    -
    \tilde A_e \, H_d \cdot \tilde L \, \tilde \ell_R^c
    + h.c.
    \right) \,,
\end{eqnarray}
where $\tilde M_L$, $\tilde M_{\nu_R}$ and $\tilde A_\nu$ are
supersymmetry breaking mass parameters.  We parameterize $\tilde
A_\nu$ as
\begin{eqnarray}
  \tilde A_\nu 
  = y_\nu \, A_\nu
  = y_\nu \, a_\nu \, \tilde M_L \,,
\end{eqnarray}
where $a_\nu$ is a dimension-less constant and $|a_\nu|$ is expected
to be of the order of unity in simple models of supergravity.  Due to
the smallness of the neutrino Yukawa coupling constants, the
left-right mixing of sneutrinos is very small.  Thus, we can
treat the left- and right-handed sneutrinos as mass eigenstates.
Their masses are given by
\begin{eqnarray}
  \label{eq:snumasses}
  m_{\tilde \nu_L}^2 \simeq \tilde M_{L}^2 
  + \frac 1 2 \, \cos 2 \beta \, m_Z^2 \,,~~~~
  m_{\tilde \nu_R}^2 \simeq \tilde M_{\nu_R}^2 \,,
\end{eqnarray}
where we have omitted the negligible contributions of $m_\nu^2$.
The left-right mixing angle of sneutrinos is denoted by $\Theta$,
which is found as
\begin{eqnarray}
  \label{eq:TH}
  \tan 2 \Theta =
  \frac{ 2 m_\nu |  \cot \beta \, \mu_H - A_\nu^\ast |}
  { m_{\tilde \nu_L}^2 - m_{\tilde \nu_R}^2 } \,,
\end{eqnarray}
and $\Theta$ is highly suppressed because of the smallness of $m_\nu$.
It can be seen from (\ref{eq:snumasses}) that masses of right-handed
sneutrinos are determined solely by the supersymmetry breaking masses,
and hence the lightest right-handed sneutrino may become the LSP.

%%%%%%%%%%%%%%%%%%%%%%%%%%%%%%%%%%%%%%%%%%%%%%%%%%%%%%%%%%%%%%%%%%%%
%%%%%%%%%%%%%%%%%%%%%%%%%%%%%%%%%%%%%%%%%%%%%%%%%%%%%%%%%%%%%%%%%%%%
%%%%%%%%%%%%%%%%%%%%%%%%%%%%%%%%%%%%%%%%%%%%%%%%%%%%%%%%%%%%%%%%%%%%
\section{Production of the LSP right-handed sneutrino}
\label{sec:production}
%%%%%%%%%%%%%%%%%%%%%%%%%%%%%%%%%%%%%%%%%%%%%%%%%%%%%%%%%%%%%%%%%%%%
%%%%%%%%%%%%%%%%%%%%%%%%%%%%%%%%%%%%%%%%%%%%%%%%%%%%%%%%%%%%%%%%%%%%
%%%%%%%%%%%%%%%%%%%%%%%%%%%%%%%%%%%%%%%%%%%%%%%%%%%%%%%%%%%%%%%%%%%%

As we have shown in Ref.~\cite{Asaka:2005cn}, the LSP right-handed
sneutrino is a viable candidate of CDM.  This is because it is stable
under the $R$-parity conservation, and also because it has only
suppressed interactions, {\it i.e.}, the neutrino Yukawa interactions
and the gauge interactions through the left-right mixing.

In this section, we discuss production processes of right-handed
sneutrino $\tilde\nu_R$ in the early universe and calculate the
density parameter of right-handed sneutrino:
\begin{eqnarray}
  \Omega_{\tilde \nu_R} = 
  \frac{\rho_{\tilde \nu_R, 0}}{\rho_{\rm cr}} \,,
\end{eqnarray}
where $\rho_{\tilde \nu_R, 0}$ is the present energy density of
$\tilde \nu_R$ and $\rho_{\rm cr} = 1.05 \times 10^{-5} \, h^2$
GeV/cm$^3$ is the critical density.  Importantly, there are two types
of production processes of right-handed-sneutrino LSP; one is the
decay of MSSM superparticles in the chemical equilibrium and the other
is the decay of the MSSM-LSP after freeze out.  We have considered the
former effect in Ref.~\cite{Asaka:2005cn} and found that, in most of
the cases, $\Omega_{\tilde \nu_R}$ via the decay of MSSM
superparticles in the chemical equilibrium becomes smaller than
$\Omega_{\rm DM}$.  Thus, it is probable that the relic
$\tilde{\nu}_R$ are mostly from the decay of the MSSM-LSP after
freeze-out.  So, we first consider the latter effect in detail
although we will also discuss the former production process later.

%%%%%%%%%%%%%%%%%%%%%%%%%%%%%%%%%%%%%%%%%%%%%%%%%%%%%%%%%%%%%%%%%%%%%%%%
%%%%%%%%%%%%%%%%%%%%%%%%%%%%%%%%%%%%%%%%%%%%%%%%%%%%%%%%%%%%%%%%%%%%%%%%
\subsection{$\tilde{\nu}_R$ production via MSSM-LSP decay}
\label{sec:FO}
%%%%%%%%%%%%%%%%%%%%%%%%%%%%%%%%%%%%%%%%%%%%%%%%%%%%%%%%%%%%%%%%%%%%%%%%
%%%%%%%%%%%%%%%%%%%%%%%%%%%%%%%%%%%%%%%%%%%%%%%%%%%%%%%%%%%%%%%%%%%%%%%%

Let us discuss the production of $\tilde\nu_R$ via decays of the MSSM-LSP
after freeze out.  As we will discuss, the density parameter of
$\tilde\nu_R$ is sensitive to the MSSM parameters.  In this paper, we
will not study the complete parameter space since it is beyond the
scope of our purpose; we rather illustrate the basic features using a
simple model of supersymmetry breaking.  Here we consider, as an
example, the minimal supergravity model with the gauge coupling
unification at $M_G \simeq 2 \times 10^{16}$ GeV.  This model is
described by $m_0$ (the universal scalar mass at $M_G$), $m_{1/2}$
(the unified gaugino mass at $M_G$), and $\tilde A_G$ (the universal
$A$-parameter at $M_G$) together with $\tan \beta$ and the sign of $\mu_H$.  Then, the on-shell mass
of $\tilde \nu_R$ is given by
\begin{eqnarray}
  m_{\tilde \nu_R} = m_0 \,,
\end{eqnarray}
since the renormalization group (RG) evolution of $m_{\tilde\nu_R}$
is negligible due to the smallness of $y_\nu$.

Adopting the minimal supergravity model, three right-handed sneutrinos
are almost degenerate.  The mass differences are mainly from the
effects of RG evolution through $y_\nu$, and hence they are very
small.\footnote
%%%
{ Other sources of the mass differences are the neutrino
masses and the contributions from the left-right mixing of sneutrinos.
}
%%%
Consequently, all the three right-handed sneutrinos are stable within
the age of the universe and contribute to the dark-matter density.

Since the neutrino Yukawa coupling constants are negligibly small, the
decay of the MSSM-LSP into $\tilde \nu_R$ is extremely suppressed.
Detailed value of the lifetime of the MSSM-LSP depends on MSSM
parameters.  However, as we will show later, lifetime of the MSSM-LSP
is long enough so that the decay of the MSSM-LSP occurs sufficiently
after its freeze out.

%%%%%%%%%%%%%%%%%%%%%%%%%%%%%%%%%%%%%%%%%%%%%%%%%%%%%%%%%%%%%%%%%%%%%
\begin{figure*}
    \begin{center}
        \includegraphics[scale=1.3]{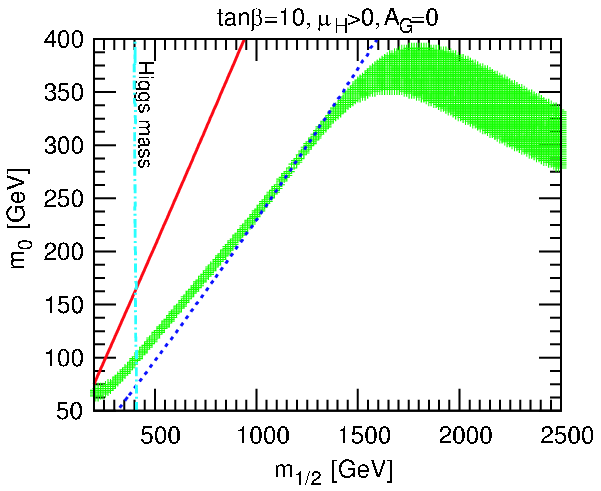}%
        \includegraphics[scale=1.3]{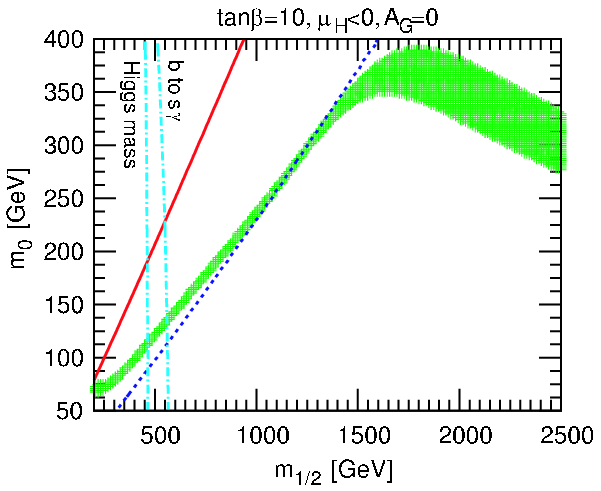}
        \includegraphics[scale=1.3]{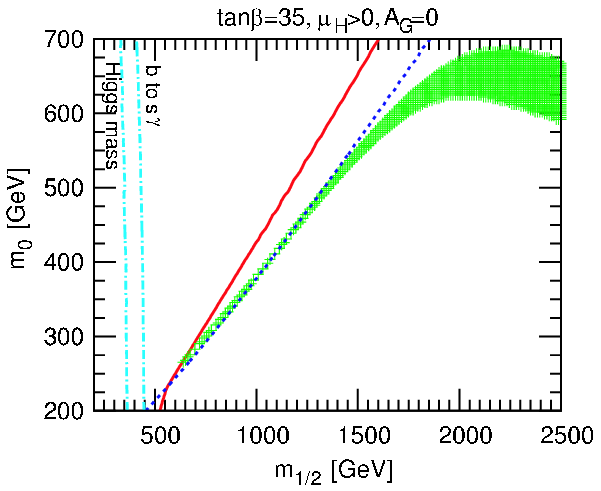}%
        \includegraphics[scale=1.3]{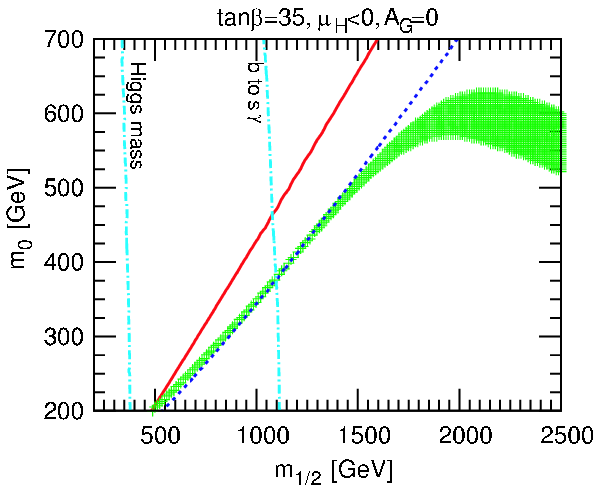}
    \end{center}
    \caption{The parameter range to realize $\tilde \nu_R$-CDM 
    in the minimal supergravity model with $A_G = 0$ (shown by green
    ``+'' marks).  The solid line represents $m_{\tilde
    \nu_R}=m_{\mbox{\tiny MSSM-LSP}}$; $\tilde \nu_R$ becomes the LSP
    below this line.  The MSSM-LSP is the lightest neutralino $\tilde
    \chi_1^0$ or the lighter stau $\tilde \tau_1$ in the region above
    or below the dashed line.  The vertical dot-dashed lines show the
    lower bounds on $m_{1/2}$ from the Higgs mass and $b \to s
    \gamma$.}
    \label{fig:CONT_MSUGRA}
\end{figure*}
%%%%%%%%%%%%%%%%%%%%%%%%%%%%%%%%%%%%%%%%%%%%%%%%%%%%%%%%%%%%%%%%%%%%%

The density parameter of $\tilde \nu_R$ from the decay of the MSSM-LSP
after freeze out is given by
\begin{eqnarray}
  \label{eq:omega_snrFO}
  \Omega_{\tilde \nu_R}^{\mbox{\tiny FO}} =
    \frac{m_{\tilde \nu_R}}{m_{\mbox{\tiny MSSM-LSP}}} \, 
    \Omega_{\mbox{\tiny MSSM-LSP}} \,,
\end{eqnarray}
where $\Omega_{\mbox{\tiny MSSM-LSP}}$ is the (would-be) present
density parameter of the MSSM-LSP for the case where it is stable.
$\Omega_{\mbox{\tiny MSSM-LSP}}$ is estimated using conventional
method since the neutrino Yukawa coupling constants are very
small.\footnote
%%%%
{ In this analysis, we assume that only the thermal relic of the
MSSM-LSP contributes to $\Omega_{\mbox{\tiny MSSM-LSP}}$.  The
MSSM-LSP might be also produced via non-thermal processes
\cite{Moroi:1999zb}.  Even in this case, if $\Omega_{\mbox{\tiny
MSSM-LSP}}$ is given as (\ref{eq:omega_mssm}), the $\tilde \nu_R$-CDM
is realized.  }
%%%%
$\tilde \nu_R$ can account for the present dark-matter density if
$\Omega_{\tilde \nu_R}^{\mbox{\tiny FO}} = \Omega_{\mbox{\tiny DM}}$.
We note here that $\Omega_{\tilde \nu_R}^{\mbox{\tiny FO}}$ is
insensitive to the reheating temperature $T_R$ after inflation as long
as $T_R$ is higher than $T_F$, where $T_F$ is the freeze-out
temperature which is roughly given by $T_F \sim m_{\mbox{\tiny
MSSM-LSP}}/20$.

Based on Eq.\ (\ref{eq:omega_snrFO}), $\tilde{\nu}_R$-CDM is realized
with some choice of MSSM parameters which satisfies the following
relation:
\begin{eqnarray}
  \label{eq:omega_mssm}
    \Omega_{\mbox{\tiny MSSM-LSP}} = 
    \frac{m_{\mbox{\tiny MSSM-LSP}}}{m_{\tilde \nu_R}} \, 
    \Omega_{\mbox{\tiny DM}} \,.
\end{eqnarray}
Since $m_{\mbox{\tiny MSSM-LSP}}\neq m_{\tilde \nu_R}$, this implies
that the MSSM parameters realizing $\tilde{\nu}_R$-CDM are different
from those for the conventional scenario where the MSSM-LSP, say the
lightest neutralino, becomes CDM.  Furthermore, $\tilde{\nu}_R$-CDM is
possible even if the MSSM-LSP is an electrically charged and/or
colored superparticle, and hence the $\tilde{\nu}_R$-CDM is realized
in a wider parameter range compared with the MSSM-LSP dark matter.

In the parameter range of interest, the MSSM-LSP is the lightest
neutralino $\tilde \chi_1^0$ which is Bino-like or the lighter stau
$\tilde \tau_1$.  We estimate their (would-be) relic density
$\Omega_{\mbox{\tiny MSSM-LSP}} h^2$ by using the micrOMEGAs
package~\cite{Belanger:2001fz}, and then calculate $\Omega_{\tilde
\nu_R}^{\mbox{\tiny FO}} h^2$ using Eq.~(\ref{eq:omega_snrFO}).  In
Fig.~\ref{fig:CONT_MSUGRA}, we show the parameter region where
$\Omega_{\tilde \nu_R}^{\mbox{\tiny FO}} h^2$ is in the range
(\ref{eq:omega_dm}).  Here, for simplicity, we set $A_G=0$.  In the
figure, we also impose the following phenomenological constraints; the
mass of the lighter neutral Higgs boson (the standard-model like Higgs
boson) be larger than the experimental bound \cite{Yao:2006px}
\begin{eqnarray}
  \label{eq:Higgsmassbound}
  m_{h^0} > 114.4 \GeV \,,
\end{eqnarray}
and the branching ratio of $b \to s \gamma$ be in 
the three-sigma range of the observational data~\cite{unknown:2006bi}:
\begin{eqnarray}
    \mbox{Br} ( b \to s \gamma ) =
    (3.55 \pm 0.78) \times 10^{-4} \,.
\end{eqnarray}

One can see that $\tilde{\nu}_R$ can be CDM of the universe in a wide
parameter region.  In addition, we emphasize that $\tilde{\nu}_R$-CDM
can be realized not only in the case where $\tilde\chi_1^0$ is the
MSSM-LSP but also in the case with $\tilde\tau_1$-MSSM-LSP.  This is
an interesting feature of the present scenario.  We also stress here
that $\Omega_{\tilde\nu_R}^{\mbox{\tiny FO}}$ is insensitive to the
neutrino Yukawa coupling constants.

%%%%%%%%%%%%%%%%%%%%%%%%%%%%%%%%%%%%%%%%%%%%%%%%%%%%%%%%%%%%%%%%%%%%%
\begin{figure}[tb]
    \begin{center}
        \includegraphics[scale=1.3]{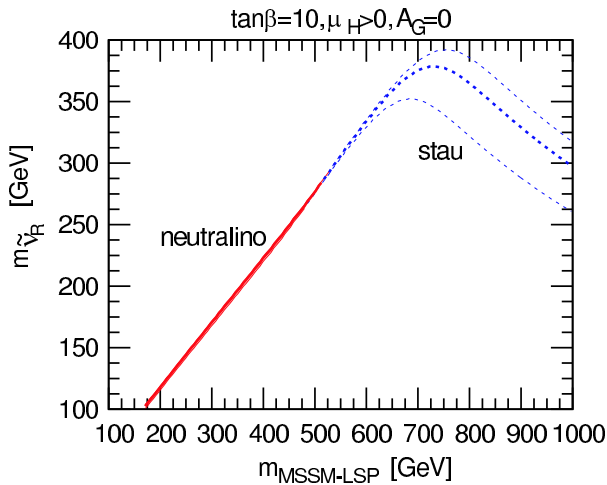}
    \end{center}
    \caption{ The mass of $\tilde{\nu}_R$-CDM in the minimal
    supergravity model. We take here $A_G=0$, $\tan \beta = 10$ and
    $\mu_H >0$.  The thick line corresponds to $\Omega_{\tilde
    \nu_R}^{\mbox{\tiny FO}} h^2 = 0.105$, and the thin lines
    correspond to $\Omega_{\tilde \nu_R}^{\mbox{\tiny FO}} h^2$ in the
    range (\ref{eq:omega_dm}).  The solid or dashed lines are
    for the $\tilde \chi_1^0$ or $\tilde
    \tau_1$ MSSM-LSP, respectively.  }
    \label{fig:M_MSSMLSP_tb10mP}
\end{figure}
%%%%%%%%%%%%%%%%%%%%%%%%%%%%%%%%%%%%%%%%%%%%%%%%%%%%%%%%%%%%%%%%%%%%%

In Fig.\ \ref{fig:M_MSSMLSP_tb10mP}, we show the mass of right-handed
sneutrino which gives the correct value of $\Omega_{\tilde
  \nu_R}^{\mbox{\tiny FO}}$ to realize $\tilde{\nu}_R$-CDM as a
function of the mass of the MSSM-LSP.  Here, we take $A_G=0$,
$\tan\beta =10$ and $\mu_H>0$; in this case, when $\Omega_{\tilde
  \nu_R}^{\mbox{\tiny FO}} h^2 = 0.105$, the MSSM-LSP is the lightest
neutralino for $m_{\mbox{\tiny MSSM-LSP}}\lesssim 510\ {\rm GeV}$,
and is stau for $m_{\mbox{\tiny MSSM-LSP}}\gtrsim 510\ {\rm GeV}$.
Fig.\ \ref{fig:M_MSSMLSP_tb10mP} shows that the mass of the
right-handed sneutrino is bounded from above to realize
$\tilde{\nu}_R$-CDM.

In the region where $\tilde \chi_1^0$ becomes the MSSM-LSP, we find an
approximate relation $m_{\tilde \chi_1^0} \simeq 1.8 \, m_{\tilde
\nu_R}$ to realize $\Omega_{\tilde \nu_R}^{\mbox{\tiny FO}} =
\Omega_{\mbox{\tiny DM}}$.  In this case, the mass of the
$\tilde{\nu}_R$-CDM is well determined from $m_{\tilde \chi_1^0}$ even
if one varies $\Omega_{\tilde \nu_R}^{\mbox{\tiny FO}} h^2$ in the
range given in Eq.\ (\ref{eq:omega_dm}).  Moreover, we find that
masses of $\tilde \chi_1^0$ and $\tilde \tau_1$ are close to each
other and their co-annihilation effects is important in estimating
$\Omega_{\mbox{\tiny MSSM-LSP}}$.  (See the following discussion and
Fig.~\ref{fig:Mstau_tb10mP}.)

As $m_{\tilde \chi_1^0}$ becomes larger, the MSSM-LSP is $\tilde
\tau_1$.  Then, the relation $\Omega_{\tilde \nu_R}^{\mbox{\tiny FO}}
= \Omega_{\mbox{\tiny DM}}$ is realized even when $m_{\tilde
\chi_1^0}$ is much larger than $\sim 1$\ TeV.  For $m_{\tilde
\chi_1^0} \gg 700$ GeV, the yield variable of $\tilde \tau_1$, which
is defined by the ratio of the number density to the entropy density,
is given by~\cite{Asaka:2000zh}
\begin{eqnarray}
  Y_{\tilde \tau_1} \simeq 10^{-12} \times c_Y 
  \left( \frac{m_{\tilde \tau_1}}{1\TeV} \right)\,,
\end{eqnarray}
where $c_Y$ is a constant of the order of unity; we numerically found
that $c_Y = 1.2-1.3$ in the parameter region we are interested in.
We can then obtain the relic density of $\tilde \nu_R$ from $\tilde
\tau_1$ decay as
\begin{eqnarray}
  \label{eq:omega_stau}
  \Omega_{\tilde \nu_R}^{\mbox{\tiny FO}} h^2
  \simeq 0.027 \times c_Y 
  \left( \frac{ m_{\tilde \tau_1} }{1\TeV} \right) 
  \left( \frac{ m_{\tilde \nu_R} }{100\GeV} \right) \,.
\end{eqnarray}
The numerical calculation shows that, when $m_{\tilde \nu_R} = m_0 =
100$ GeV, for example, $\Omega_{\tilde \nu_R}^{\mbox{\tiny FO}} h^2 =
0.105$ is realized for $m_{\tilde \tau_1} \simeq 3.1$ TeV, which
corresponds to $m_{1/2} \simeq 8.7$ TeV and $m_{\tilde \chi_1^0}
\simeq 4.1$ TeV.  Even when we take larger value of $m_{\tilde
\tau_1}$ (and $m_{\tilde \chi_1^0}$), the $\tilde{\nu}_R$-CDM is
possible by taking $m_{\tilde \nu_R} \lesssim 100$ GeV, as seen from
Eq.\ (\ref{eq:omega_stau}).

Now, we come to a position to discuss the lifetime of the MSSM-LSP.
As we mentioned, Eq.\ (\ref{eq:omega_snrFO}) is applicable when the
MSSM-LSPs decay after the time of freeze out.  In order to see if this
is the case, let us estimate the decay rate of the MSSM-LSP.  When the
MSSM-LSP is the Bino-like neutralino, it decays into $\tilde \nu_R$
and anti-neutrino (and also its CP conjugate state) through the
left-right mixing of sneutrinos.  The decay rate is estimated as
\begin{eqnarray}
    \Gamma_{\rm \tilde \chi_1^0} 
    \eqn{\simeq} 
    \Gamma_{\tilde B \to \tilde \nu_R \bar \nu}
    +
    \Gamma_{\tilde B \to \tilde \nu_R^c \nu}
    \nonumber \\
    \eqn{=} 
    \frac{g_1^2 \, \Theta^2 }{32 \, \pi} \,  
    m_{\tilde \chi_1^0 }
    \left( 1 - \frac{m_{\tilde \nu_R}^2}{m_{\tilde \chi_1^0}^2} \right)^2
    \,,
\end{eqnarray}
where $g_1$ is a gauge coupling constant of U(1)$_Y$.  Notice that the
Bino-like neutralino $\tilde \chi_1^0$ universally couples to
(s)neutrinos in all three generations.  Thus, the decay process into
the sneutrino which interacts via the strongest Yukawa interaction is
the most important since the mixing angle $\Theta$ is proportional to
the Yukawa coupling constant.

When the MSSM-LSP is $\tilde \tau_1$, the situation is slightly more
complicated since the dominant decay mode depends on the mass
difference between $\tilde \tau_1$ and $\tilde{\nu}$.  When $m_{\tilde
\tau_1}>m_{\tilde \nu_R}+m_W$, the decay rate is given by
\begin{eqnarray}
  \Gamma_{\tilde \tau_1} 
  \eqn{\simeq}
  \Gamma_{\tilde \tau_1 \to \tilde \nu_R W}
  \nonumber \\
  \eqn{=}
  \frac{ g_2^2 \, \Theta^2 }{32 \pi}
  |U^{(\tilde \tau_1)}_{L1}|^2 \,
  |U^{\rm MNS}_{\tau 3}|^2
  \frac{m_{\tilde \tau_1}^3}{m_W^2} 
  \nonumber \\
  \eqn{} \times
  \left[
    1 - \frac{ 2 ( m_{\tilde \nu_R}^2 + m_{W}^2 ) }{ m_{\tilde \tau_1}^2 }
      + \frac{ ( m_{\tilde \nu_R}^2 - m_{W}^2 )^2 }{ m_{\tilde \tau_1}^4 }
  \right]^{3/2} ,~~~
\end{eqnarray}
where $g_2$ is a gauge coupling constant of SU(2)$_L$, $m_W$ is the
$W$-boson mass, and $U^{(\tilde \tau_1)}$ is the mixing matrix of
staus which relates the gauge eigenstates and mass eigenstates as
\begin{eqnarray}
  \left( 
    \begin{array}{c}
      \tilde \tau_L \\
      \tilde \tau_R
    \end{array}
  \right)
  =
  U^{(\tilde \tau)}
  \left( 
    \begin{array}{c}
      \tilde \tau_1 \\
      \tilde \tau_2
    \end{array}
  \right) 
\,,
\end{eqnarray}
where $\tilde \tau_1$ and $\tilde \tau_2$ are mass eigenstates with
$m_{\tilde \tau_1} \le m_{\tilde \tau_2}$.  In addition, $U^{\rm MNS}$
is the neutrino mixing matrix.  Here, we use $U^{\rm MNS}_{\tau
3}=1/\sqrt{2}$.  Notice that the above expression is relevant for the
case with hierarchical neutrino mass; for the degenerate case, decay
processes into all three sneutrinos may be equally important.  For
example, when the mass differences among three left-handed sleptons
are negligible, $|U^{\rm MNS}_{\tau 3}|^2$ should be replaced by $1$.
Note that $\tilde \tau_1$ may also decay into $\tilde \nu_R$ and the
charged Higgs boson.  Such a decay channel is, however, kinematically
forbidden in the parameter region which we are interested in.  When
the decay process $\tilde\tau_1\to\tilde\nu_RW$ is kinematically
forbidden, $\tilde\tau_1$ mainly decays as $\tilde \tau_1 \to \tilde
\nu_R \, \ell \, \overline \nu$, $\tilde \nu_R \, q \, \overline q'$
with exchanging virtual $W$-boson.

%%%%%%%%%%%%%%%%%%%%%%%%%%%%%%%%%%%%%%%%%%%%%%%%%%%%%%%%%%%%%%%%%%%%%
\begin{figure}[tb]
 \begin{center}
     \includegraphics[scale=1.3]{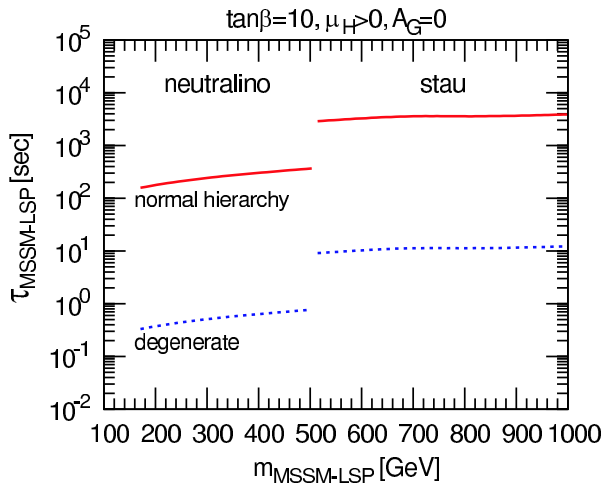}%
 \end{center}
 \caption{ The lifetime of the MSSM-LSP yielding $\Omega_{\tilde
     \nu_R}^{\mbox{\tiny FO}} h^2 = 0.105$ in the minimal supergravity
   model.  We take here $A_G=0$, $\tan \beta =10$ and $\mu_H >0$.  
   The MSSM-LSP is $\tilde \chi_1^0$ or $\tilde \tau_1$ 
   when $m_{\rm MSSM-LSP}$ is smaller or larger than about 510 GeV.
   The solid lines are the lifetime assuming the normal
   hierarchy of neutrino masses and taking $m_\nu = \sqrt{[\Delta
     m_\nu^2]_{\mbox{\tiny atom}}}$, 
   while the dashed lines are the lifetime
   assuming the degenerate neutrino masses and taking
   $m_\nu = 0.67$ eV.}
 \label{fig:LIFE_MSSMLSP}
\end{figure}
%%%%%%%%%%%%%%%%%%%%%%%%%%%%%%%%%%%%%%%%%%%%%%%%%%%%%%%%%%%%%%%%%%%%%

Using $y_{\nu}\simeq 3.0\times 10^{-13}$, which is the suggested value
of the neutrino Yukawa coupling constant in the case with the normal
hierarchy neutrino mass matrix, the lifetime of the MSSM-LSP for the
case yielding $\Omega_{\tilde \nu_R}^{\mbox{\tiny FO}} h^2=0.105$ is
shown in Fig.~\ref{fig:LIFE_MSSMLSP}.  Here, we take $A_G=0$ to
illustrate the typical value of the lifetime.  We can see that the
lifetime is roughly of the order of $10^2$--$10^3$ sec, and hence the
decay of the MSSM-LSP occurs after its freeze-out time.  When neutrino
masses are degenerate, the lifetime of the MSSM-LSP becomes
significantly shorter because the neutrino Yukawa coupling constants
are enhanced and also because the MSSM-LSP decays into three (almost)
degenerate right-handed sneutrinos.  Even in this case, the MSSM-LSP
decays into $\tilde \nu_R$ after the time of the freeze-out.
Therefore, calculation with Eq.\ (\ref{eq:omega_snrFO}) is justified
with any of the neutrino mass matrix.

Here, we should comment on the fact that the lifetime of the MSSM-LSP
depends on the $A$-parameter since the left-right mixing angle of the
neutrino $\Theta$ is sensitive to it.  By solving RG equations, we
found that $A_\nu$ parameter at the scale $m_Z$ is given by
\begin{eqnarray}
  A_\nu \simeq A_G - 0.59 \, m_{1/2}.
  \label{A_nu(Mz)}
\end{eqnarray}
Thus, as $A_G$ changes, the lifetime of the MSSM-LSP varies.

We would like to note that the $\tilde{\nu}_R$-CDM scenario and
MSSM-LSP dark matter scenario are realized in different parameter
space.  In particular, we emphasize that $\Omega_{\mbox{\tiny
MSSM-LSP}}>\Omega_{\mbox{\tiny DM}}$ is required if the right-handed
sneutrino produced from the decay of MSSM-LSP after freeze out becomes
CDM.  Using this fact, we can potentially distinguish the scenario of
$\tilde{\nu}_R$-CDM and MSSM-LSP dark matter with precise
determinations of the properties of MSSM particles by future collider
experiments.  To see this, in Fig.~\ref{fig:Mstau_tb10mP}, we show the
regions on $m_{\tilde \chi_1^0}$ vs.\ $m_{\tilde \tau_1}-m_{\tilde
\chi_1^0}$ plane where the density parameter of the LSP becomes
consistent with the presently observed dark matter density for the
cases where the LSP is $\tilde{\nu}_R$ and $\tilde\chi_1^0$.  We can
see that, for a given value of $m_{\tilde \chi_1^0}$, the suggested
values of $m_{\tilde \tau_1}$ are different by 5 GeV or so between two
scenarios.  We also show the correlations between $m_{\tilde \tau_1}$
and $m_{\tilde \chi_1^+}$ (with $\tilde \chi_1^+$ being the lighter
chargino) in two scenarios.  We can see that, for a given $m_{\tilde
\tau_1}$, the suggested value of $m_{\tilde \chi_1^+}$ is different by
$\sim 10$ GeV.  Thus, with the precise measurements of the masses of
MSSM superparticles by future collider experiments, like the LHC and
ILC, we may be able to distinguish the $\tilde{\nu}_R$-CDM and the
MSSM-LSP-CDM scenarios.

%%%%%%%%%%%%%%%%%%%%%%%%%%%%%%%%%%%%%%%%%%%%%%%%%%%%%%%%%%%%%%%%%%%%%
\begin{figure}[tb]
 \begin{center}
     \includegraphics[scale=1.3]{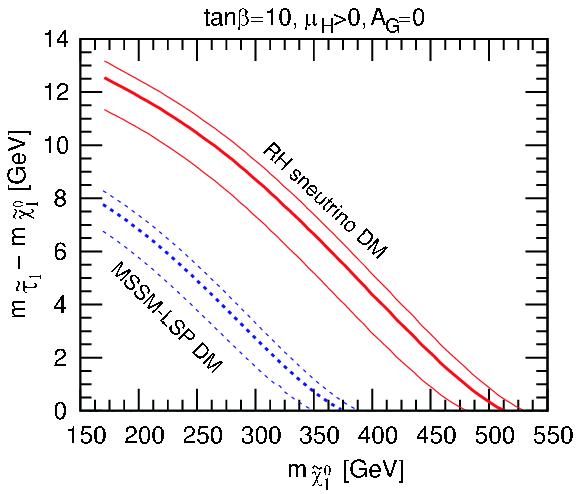}
     \includegraphics[scale=1.3]{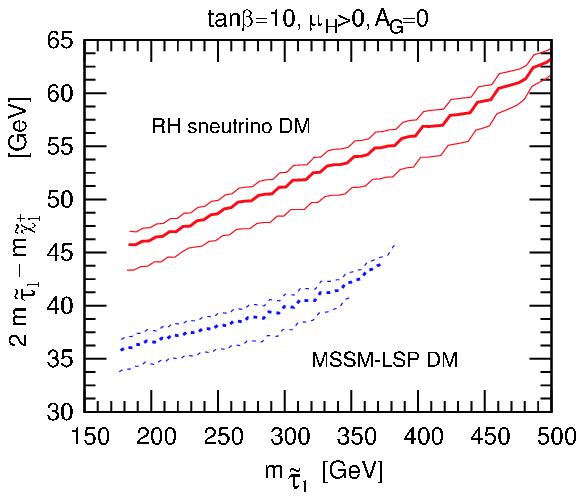}
 \end{center}
 \caption{ The mass differences $m_{\tilde \tau_1} - m_{\tilde
     \chi_1^0}$ and $2m_{\tilde \tau_1}-m_{\tilde \chi_1^+}$ realizing
   $\tilde \nu_R$-CDM in the minimal supergravity model.  We take
   here $A_G = 0$, $\tan \beta = 10$ and $\mu_H >0$.  The thick
   solid line corresponds to $\Omega_{\tilde \nu_R}^{\mbox{\tiny FO}} h^2 
   = 0.105$ and the thin solid lines 
   correspond to $\Omega_{\tilde \nu_R}^{\mbox{\tiny FO}} h^2$ 
   in the range (\ref{eq:omega_dm}).
   For comparison, we also show the mass differences in the
   case where the MSSM-LSP $\tilde \chi_1^0$ becomes dark matter by
   dashed lines.}
 \label{fig:Mstau_tb10mP}
\end{figure}
%%%%%%%%%%%%%%%%%%%%%%%%%%%%%%%%%%%%%%%%%%%%%%%%%%%%%%%%%%%%%%%%%%%%%

\subsection{Constraints from BBN}

As one can see from Fig.~\ref{fig:LIFE_MSSMLSP}, the lifetime of the
MSSM-LSP may be so long that the decay of the MSSM-LSP occurs around
or after the epoch of BBN.  If so, abundances of light elements which
are produced by the standard BBN reactions are affected by the decay
of the MSSM-LSP and, consequently, the success of BBN may be spoiled.

Effects of late-decaying particles on BBN are intensively studied in
Ref.~\cite{BBN_KKM}; the most important effects of the late-decaying
particles are from $p\leftrightarrow n$ conversion,
hadro-dissociation, and/or photo-dissociation process, depending on
the lifetime.  For the case of our interest, lifetime of the MSSM-LSP
is $\lesssim 10^{4}\ {\rm sec}$.  In such a case, $p\leftrightarrow n$
conversion and hadro-dissociation are important.  If the abundance of
the MSSM-LSP is too large, the abundances of light elements are too
much affected to be consistent with observations by these processes.
In order not to spoil the success of BBN, upper bound on the
combination $B_{\rm had}E_{\rm vis}Y_{\mbox{\tiny MSSM-LSP}}$ is
obtained.  In our case, $B_{\rm had}$ is the hadronic branching ratio
of the MSSM-LSP, $E_{\rm vis}$ is the net energy carried away by
hadrons in the decay of the MSSM-LSP, and $Y_{\mbox{\tiny MSSM-LSP}}$
is the yield of the MSSM-LSP.  In order to see if $\tilde{\nu}_R$-CDM
is viable, we will calculate these quantities in the following.

%%%%%%%%%%%%%%%%%%%%%%%%%%%%%%%%%%%%%%%%%%%%%%%%%%%%%%%%%%%%%%%%%%%%%
\begin{figure}[tb]
 \begin{center}
   \includegraphics[scale=1.3]{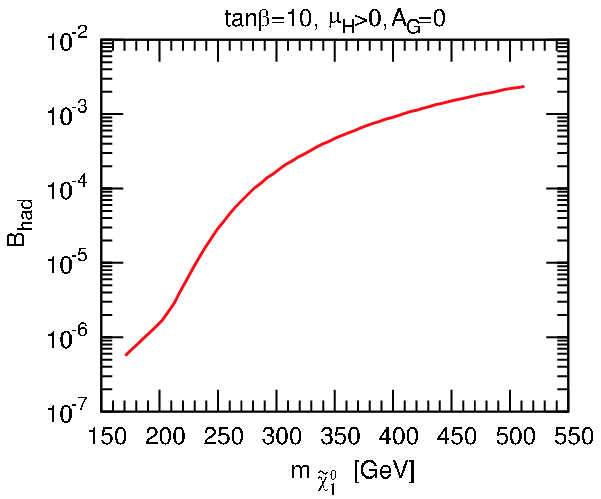}
   \includegraphics[scale=1.3]{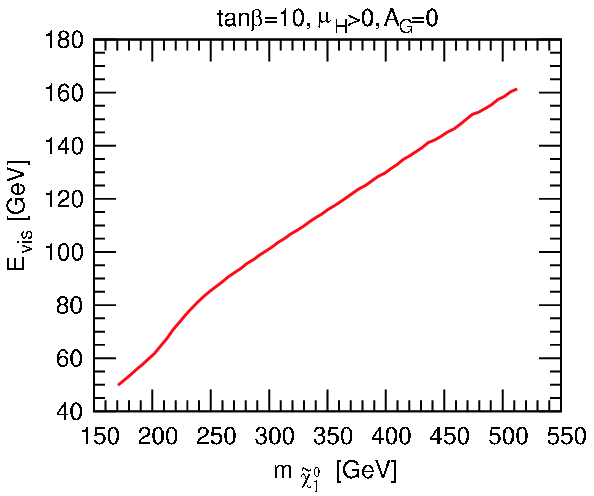}
 \end{center}
 \caption{The hadronic branching ratio $B_{\rm had}$
   and the net energy carried by hadrons $E_{\rm vis}$
   in the $\tilde \chi_1^0$ MSSM-LSP decay.
   Here we consider the minimal supergravity model
   with $A_G=0$, $\tan \beta = 10$ and $\mu_H >0$,
   and require $\Omega_{\tilde \nu_R}^{\rm FO}h ^2 = 0.105$.}
 \label{fig:BHADchi0}
\end{figure}
%%%%%%%%%%%%%%%%%%%%%%%%%%%%%%%%%%%%%%%%%%%%%%%%%%%%%%%%%%%%%%%%%%%%%

We first consider the case where the MSSM-LSP is $\tilde \chi_1^0$.
Even though the dominant decay mode of $\tilde \chi_1^0$ is into
neutrino and right-handed sneutrino, there still exist hadronic decay
modes:
\begin{eqnarray}
  \tilde \chi_1^0 \to 
  \tilde \nu_R \, \overline{\nu} \, q \, \overline{q} \,,~
  \tilde \nu_R^\ast \, \nu \, q \, \overline{q} \,,~
  \tilde \nu_R \, {\ell^+} \,  q \, \overline{q}' \,,~
  \tilde \nu_R^\ast \, {\ell^-} \,  q \, \overline{q}' \,,~~~
\end{eqnarray}
where $q$ and $q'$ are quarks and $\ell^{\pm}$ are charged leptons.
These processes are mediated by on-shell and/or off-shell gauge
bosons.  We have numerically calculated the decay rate of these
processes and obtained $B_{\rm had}$ and $E_{\rm vis}$.  Results for
the case with $A_G=0$, $\tan \beta = 10$ and $\mu_H >0$ are shown in
Fig.~\ref{fig:BHADchi0}.  As one can see, hadronic branching ratio
becomes $O(10^{-4}-10^{-3})$ when the mass of $\tilde\chi_1^0$ is
relatively large.  This is because, when $m_{\tilde\chi_1^0}$ is large
enough, the decay modes $\tilde\chi_1^0\rightarrow\nu_Rl^{+}W^{-}$ and
$\tilde\chi_1^0\rightarrow\nu_R\bar{\nu}Z$ (and their CP conjugated
processes) are kinematically allowed and hence hadrons can be produced
via the ``three-body'' decay processes.  On the contrary, when
$m_{\tilde\chi_1^0}$ is small, the weak bosons are always off-shell
and $B_{\rm had}$ is suppressed.

%%%%%%%%%%%%%%%%%%%%%%%%%%%%%%%%%%%%%%%%%%%%%%%%%%%%%%%%%%%%%%%%%%%%%
\begin{figure}[tb]
  \begin{center}
    \includegraphics[scale=1.3]{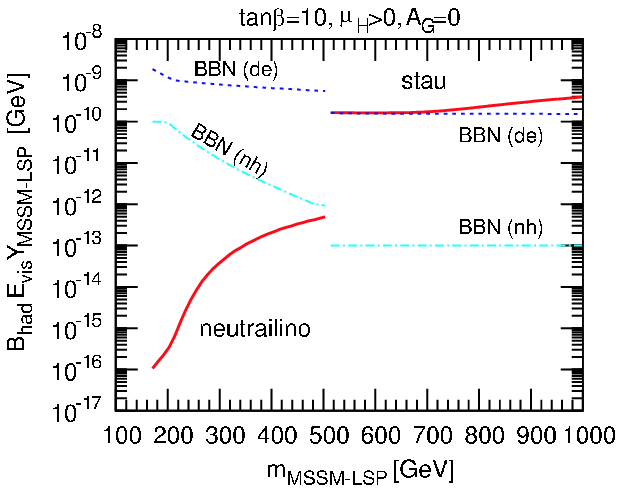}
    \includegraphics[scale=1.3]{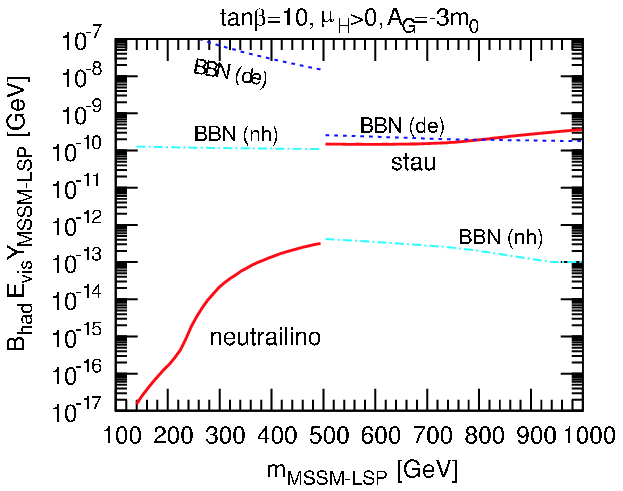}
  \end{center}
  \vspace{-0.5cm}
  \caption{ The product $B_{\rm had}E_{\rm vis}Y_{\mbox{\tiny
        MSSM-LSP}}$ in terms of the MSSM-LSP mass in the minimal
    supergravity model.  (shown by the solid lines).  The upper or
    lower panel is for the case when $A_G = 0$ or $A_G=-3 m_0$,
    respectively.  Here we take $\tan \beta = 10$ and $\mu_H >0$,
    and require $\Omega_{\tilde \nu_R}^{\mbox{\tiny FO}} h^2 =
    0.105$.  The MSSM-LSP is $\tilde \chi_1^0$ or $\tilde \tau_1$
    for $m_{\rm MSSM-LSP}$ is greater or smaller than about 510 GeV.
    The dot-dashed lines show the upper bound from the BBN
    constraint on the MSSM-LSP decay by assuming the normal
    hierarchy of neutrino masses and taking $m_\nu = \sqrt{[\Delta
        m_\nu^2]_{\mbox{\tiny atom}}}$.  The dashed lines show the
    bound by assuming the degenerate neutrino masses and taking
    $m_\nu = 0.67$ eV.}
  \label{fig:BBNCONST_tb10mP_Am0}
\end{figure}
%%%%%%%%%%%%%%%%%%%%%%%%%%%%%%%%%%%%%%%%%%%%%%%%%%%%%%%%%%%%%%%%%%%%%

For MSSM parameters which give correct value of
$\Omega_{\tilde{\nu}_R}^{\mbox{\tiny FO}}$ to realize
$\tilde{\nu}_R$-CDM, we calculate the product $B_{\rm had}E_{\rm
vis}Y_{\mbox{\tiny MSSM-LSP}}$ as a function of the MSSM-LSP mass.
The results are shown in Fig.~\ref{fig:BBNCONST_tb10mP_Am0}.  For the
same parameter set, we also calculate the lifetime of the MSSM-LSP.
Once the lifetime of the MSSM-LSP is given, we can estimate upper
bound on the product $B_{\rm had}E_{\rm vis}Y_{\mbox{\tiny MSSM-LSP}}$
in order not spoil the success of BBN.  When the lifetime is longer
than $\sim 100\ {\rm sec}$, the most important bounds come from the
abundances of D/H and $^6$Li/H; here we adopt constraints, which are
conservative ones, based on Fig.~39 in Ref.~\cite{BBN_KKM}:
\begin{eqnarray}
  \label{eq:BBN_DH}
  y < 
  \left\{
    \begin{array}{l l}
      \displaystyle
      - 13 & ~~~ \mbox{for}~~12 > x > 3.1
      \\
      \displaystyle
      - 1.8 \, x - 7.5 & ~~~\mbox{for}~~3.1 >x > 2.6
      \\
      \displaystyle
      - 6.6 \, x + 4.8 & ~~~\mbox{for}~~2.6 > x > 1.6
    \end{array}
  \right. \,,
\end{eqnarray}
where $x$ and $y$ are
\begin{eqnarray}
  x \eqn{=} \log_{10} (\tau_{\mbox{\tiny MSSM-LSP}}/\sec ) \,,
  \nonumber \\
  y \eqn{=} \log_{10} (B_{\rm had} E_{\rm vis} Y_{\mbox{\tiny
    MSSM-LSP}}/\GeV) \,.
\end{eqnarray}
With shorter lifetime, overproduction of ${\rm ^4He}$ gives an upper bound
on the abundance of the MSSM-LSP; here we show the constraint based on 
the observational constraint given in Ref.~\cite{Izotov:2003xn}:
\begin{eqnarray}
  \label{eq:BBN_Yp}
  y < 
  \left\{
    \begin{array}{l l}
      \displaystyle
      - 0.17 \, x - 9.6 & ~~~\mbox{for}~~ 3.5 > x > 0.39
      \\
      \displaystyle
      - 0.88 \, x - 9.4 & ~~~\mbox{for}~~ 0.39 > x > - 0.41
      \\
      \displaystyle
      - 3.8 \, x - 10.6 & ~~~\mbox{for}~~- 0.41 > x > - 1.2
    \end{array}
  \right. \,.
\end{eqnarray}
Upper bound on $B_{\rm had}E_{\rm vis}Y_{\mbox{\tiny MSSM-LSP}}$ is
also shown in Fig.~\ref{fig:BBNCONST_tb10mP_Am0}.  Notice that the
lifetime of MSSM-LSP depends on the MSSM parameters and also on the
neutrino Yukawa coupling constants, so we show the upper bound for
several cases.

With hierarchical neutrino mass matrix, we can see that
$\tilde{\nu}_R$-CDM is viable in most of the cases, although some
parameter region with large $m_{\tilde \chi_1^0}$ may be constrained
from the BBN, in particular for the case with $A_G = 0$.  If we
consider the degenerate case with $m_\nu = 0.67$ eV, lifetime of
$\tilde \chi_1^0$ becomes much shorter and hence the bound becomes
weaker.  Therefore, $\tilde{\nu}_R$-CDM does not conflict with the BBN
constraints irrespective of the value of $A_G$, as long as there is no
cancellation in the low-energy value of $A_\nu$ (see Eq.\
(\ref{A_nu(Mz)})).

Next, we consider the case where the MSSM-LSP is $\tilde \tau_1$.  As
mentioned before, when $\tilde \tau_1$ is the MSSM-LSP,
$\tilde{\nu}_R$-CDM is realized when $m_{\tilde \tau_1}$ is relatively
large.  In this case, $\tilde \tau_1$ mainly decays into $\tilde
\nu_R$ and $W$.  Since $W$-boson decays into quarks with the branching
ratio of $2/3$, we estimate the hadronic branching ratio of $\tilde
\tau_1$ as
\begin{eqnarray}
  B_{\rm had} = \frac{2}{3} \,.
\end{eqnarray}
In addition, $E_{\rm vis}$ is given by
\begin{eqnarray}
  E_{\rm vis}
  = \frac{ m_{\tilde \tau_1}^2 + m_W^2 - m_{\tilde \nu_R}^2 }
  { 2 m_{\tilde \tau_1} } \,.
\end{eqnarray}

In Fig.~\ref{fig:BBNCONST_tb10mP_Am0}, we also plot $B_{\rm had}E_{\rm
vis}Y_{\mbox{\tiny MSSM-LSP}}$ as a function of the mass of $\tilde
\tau_1$.  Compared to the case with $\tilde \chi_1^0$-MSSM-LSP,
$B_{\rm had}E_{\rm vis}Y_{\mbox{\tiny MSSM-LSP}}$ becomes much larger
since the hadronic branching ratio is close to $1$ in this case.  In
particular, with the normal hierarchy of the neutrino masses, lifetime
of $\tilde \tau_1$ is of the order of $10^3\ {\rm sec}$ or longer as
far as $|a_\nu|\sim O(1)$.  In this case, the whole region of the
$\tilde \tau_1$ MSSM-LSP yielding $\Omega_{\tilde \nu_R}^{\mbox{\tiny
FO}} = \Omega_{\mbox{\tiny DM}}$ is excluded.  Notice that, although
we have shown only the results when $m_{\tilde \tau_1} \lesssim 1$
TeV, the $\tilde \tau_1$ MSSM-LSP with $m_{\tilde \tau_1} \gg 1$ TeV
is also excluded since the lifetime and $E_{\rm vis}$ become larger.

On the other hand, for the case of degenerate neutrino masses, there
appear the regions in which the $\tilde \tau_1$ MSSM-LSP is consistent
with the BBN.  In such regions, the lifetime of $\tilde \tau_1$ is
sufficiently short and it only receives the weaker constraint
(\ref{eq:BBN_Yp}).  We find that the possible masses of $\tilde
\tau_1$ are relatively small and $m_{\tilde \tau_1} \gtrsim 1$ TeV is
excluded by the BBN even with degenerate neutrinos.

We also find that the light $\tilde \tau_1$ is also excluded by the
BBN when $\tan\beta$ is large.  This is because, in such a case, the
mass difference between $\tilde \nu_R$ and $\tilde \tau_1$ is smaller
than $m_W$, and the decay channel $\tilde \tau_1 \to \tilde \nu_R + W$
is kinematically blocked. If this is the case, the lifetime becomes
longer and more stringent constraints apply.

From Fig.~\ref{fig:BBNCONST_tb10mP_Am0}, we can learn that BBN
constraint puts the upper bound on the mass of the MSSM-LSP.
Interestingly, this bound excludes the possibility of the MSSM-LSP
being very heavy if $\tilde{\nu}_R$ is CDM, say being heavier than of
the order of TeV.

Before closing this subsection, we should comment that some of the
results obtained in the previous and this subsections are from the
fact that we use the minimal supergravity model.  In particular, in
the minimal supergravity model, masses of right-handed sneutrinos are
equal to $m_0$.  In this case, we obtained upper bound on the mass of
the MSSM-LSP if $\tilde{\nu}_R$ is CDM.  This result may not hold if
we consider other types of models of supersymmetry breaking.  In
particular, if we adopt smaller sneutrino masses, we can push up the
mass of the MSSM-LSP.  Furthermore, if the mass of the left-handed
sneutrino becomes close to $m_{\tilde \nu_R}$, the left-right mixing
angle of sneutrinos are enhanced.  (See Eq.~(\ref{eq:TH}).)  In this
case, the lifetime of the MSSM-LSP becomes shorter and, as a results,
the BBN constraint becomes weaker.  In such a situation, the $\tilde
\tau_1$ MSSM-LSP is possible in a wider parameter space.  Finally, we
should mention that the MSSM-LSP can be a superparticle other than the
Bino-like neutralino or the lighter stau.  To realize the $\tilde
\nu_R$ CDM, the MSSM-LSP may be electrically charged or colored
superparticle, if its relic abundance satisfies (\ref{eq:omega_mssm})
and if its decay is consistent with the BBN.

%%%%%%%%%%%%%%%%%%%%%%%%%%%%%%%%%%%%%%%%%%%%%%%%%%%%%%%%%%%%%%%%%%%%%%%%
%%%%%%%%%%%%%%%%%%%%%%%%%%%%%%%%%%%%%%%%%%%%%%%%%%%%%%%%%%%%%%%%%%%%%%%%
\subsection{Decays of superparticles in chemical equilibrium}
%%%%%%%%%%%%%%%%%%%%%%%%%%%%%%%%%%%%%%%%%%%%%%%%%%%%%%%%%%%%%%%%%%%%%%%%
%%%%%%%%%%%%%%%%%%%%%%%%%%%%%%%%%%%%%%%%%%%%%%%%%%%%%%%%%%%%%%%%%%%%%%%%

In the previous subsection, we have assumed that the $\tilde \nu_R$
production from the decay of MSSM superparticles in chemical
equilibrium is negligible.  Indeed, this is the case in most of the
parameter region.  In some case, however, such a production mechanism
also gives sizable amount of right-handed sneutrino.  (We denote its
contribution to the relic density by $\Omega_{\tilde{\nu}_R}^{\rm
CE}$.)  In this subsection, we consider $\Omega_{\tilde{\nu}_R}^{\rm
CE}$.

$\Omega_{\tilde{\nu}_R}^{\rm CE}$ is estimated by
solving the Boltzmann equation
\begin{eqnarray}
    \frac{dn_{\tilde{\nu}_R}}{dt} + 3 H n_{\tilde{\nu}_R}
    = C_{\rm decay},
    \label{BoltzmannEq}
\end{eqnarray}
where $n_{\tilde{\nu}_R}$ is the number density of $\tilde{\nu}_R$.
Denoting the distribution function of particle $x$ in the chemical
equilibrium as $f_x$, the decay term is given by
\begin{eqnarray}
  C_{\rm decay} = 
  \sum_{x,y} \int 
    \frac{d^3 k_x}{(2\pi)^3}
    \gamma_x (2 s_x + 1)
    \Gamma_{x\rightarrow \tilde{\nu}_R y}
    f_x
    \langle 1 \pm f_y \rangle_{k_x},
    \nonumber \\
\end{eqnarray}
where $\gamma_x=m_x/\sqrt{k_x^2+m_x^2}$ is the Lorentz factor,
$(2s_x+1)$ is the spin multiplicity of $x$, and $\langle 1 \pm f_y
\rangle_{k_x}$ is the averaged final-state multiplicity factor for
fixed value of initial-state momentum (with the positive and negative
signs being for bosons and fermions, respectively.)  In addition, the
summation is over all the possible production processes of
right-handed sneutrino.  Then, the relic density is found as
\begin{eqnarray}
  \Omega_{\tilde{\nu}_R}^{\rm CE} 
  = \frac{ m_{\tilde \nu_R} \, n_{\tilde \nu_R, 0}}{\rho_{\rm cr}} \,,
\end{eqnarray}
where $n_{\tilde \nu_R, 0}$ is the present value of $n_{\tilde
\nu_R}$.

Solution to the above Boltzmann equation has been already studied in
Ref.~\cite{Asaka:2005cn}; in Ref.~\cite{Asaka:2005cn}, we found that
$\Omega_{\tilde \nu_R}^{\mbox{\tiny CE}}$ becomes much smaller than
$\Omega_{\mbox{\tiny DM}}$ with neutrino mass matrix with normal
hierarchy if tri-linear scalar coupling is small.  Indeed, if
$A_\nu=0$, the relevant processes of the $\tilde{\nu}_R$ production
are the decay of Higgsinos, $\tilde{H}^0 \rightarrow \tilde{\nu}_R
\bar{\nu}_L$ and $\tilde{H}^+ \rightarrow \tilde{\nu}_R l^+_L$.  Then,
we found
\begin{eqnarray}
  \Omega_{\tilde \nu_R}^{\mbox{\tiny CE}} h^2 
  \lesssim 1.7 \times 10^{-3}
  \left( \frac{m_\nu^2}{2.8\times 10^{-3} \eV^2} \right) \,.
  \label{Omega(CE):h}
\end{eqnarray}
(Here, we assume that the neutral and charged Higgsinos are mass
eigenstates with mass $|\mu_H|$.)  The above expression is independent
of $m_{\tilde \nu_R}$, and the maximal abundance is obtained when
$|\mu_H| \simeq 2.75 m_{\tilde \nu_R}$.

It is notable that $\Omega_{\tilde{\nu}_R}^{\rm CE}$ can be as large
as $\Omega_{\rm DM}$ when the mass of left-handed sneutrino is close
to right-handed sneutrino mass.  This is because the left-right mixing
angle of sneutrinos is enhanced in such a case.  In the case of
minimal supergravity model, such a mass spectrum is hardly realized.
In general, however, left-handed and right-handed sneutrino masses are
free parameters and hence those masses may be degenerate.  Thus, in
this subsection, we work in the general framework of the MSSM rather
than adopting minimal supergravity model.  When the mass difference
between left-handed and right-handed sneutrinos are small, the
right-handed sneutrino production is dominated by the decay of Wino
and Bino in the chemical equilibrium: $\tilde{W}^0\rightarrow
\tilde{\nu}_R \bar{\nu}_L$, $\tilde{W}^+ \rightarrow \tilde{\nu}_R
l^+_L$, and $\tilde{B}\rightarrow \tilde{\nu}_R \bar{\nu}_L$ (and
their CP conjugated processes).  In the following, let us discuss this
case approximating that the mass eigenstates of charginos and
neutralinos are Wino and Bino (and Higgsinos).

The left-right mixing angle of sneutrinos in the vacuum is given in
Eq.\ (\ref{eq:TH}).  Importantly, at high temperature, the mixing
angle varies since the expectation value of the Higgs boson depends on
the temperature; we estimate the temperature-dependent mixing angle as
\begin{eqnarray}
  \label{eq:TH_T-dep}
  \Theta (T) = \frac{1}{2} \tan^{-1}
  \left(
  \frac{ 2 y_\nu v(T) |  \cot \beta \, \mu_H - A_\nu^\ast |}
  { m_{\tilde \nu_L}^2(T) - m_{\tilde \nu_R}^2 }
  \right) \,,
\end{eqnarray}
where $v(T)$ is the temperature-dependent expectation value of
standard-model-like Higgs boson $H_{\rm SM}$.  Here, we assume that
the heavier Higgses have masses much larger than $m_Z$ and that the
temperature dependence of Higgs mixing angle $\beta$ is negligible.
In addition, $m_{\tilde \nu_L}^2(T)$ is the temperature-dependent mass
of left-handed sneutrino.  With $\Theta (T)$, decay rates of the
relevant processes are given by
\begin{eqnarray}
    \Gamma_{\tilde{W}^0\rightarrow \tilde{\nu}_R \bar{\nu}_L}
    &=&
    \frac{g_2^2}{64\pi} \Theta^2 (T)
    m_{\tilde{W}} 
    \left( 1 -
        \frac{m_{\tilde{\nu}_R}^2}{m_{\tilde{W}}^2} \right)^2,
    \\ 
    \Gamma_{\tilde{W}^+ \rightarrow \tilde{\nu}_R l^+_L} 
    &=&
    \frac{g_2^2}{32\pi} \Theta^2 (T)
    m_{\tilde{W}}
    \left( 1 -
        \frac{m_{\tilde{\nu}_R}^2}{m_{\tilde{W}}^2} \right)^2,
    \label{Gamma_wino+-}
    \\
    \Gamma_{\tilde{B}\rightarrow \tilde{\nu}_R \bar{\nu}_L} 
    &=&
    \frac{g_1^2}{64\pi} \Theta^2 (T)
    m_{\tilde{B}}
    \left( 1 -
        \frac{m_{\tilde{\nu}_R}^2}{m_{\tilde{B}}^2} \right)^2.
    \label{Gamma_bino}
\end{eqnarray}

In our analysis, we approximate the Higgs potential in thermal bath as
\cite{Dine:1992wr}
\begin{eqnarray}
    V_T &\simeq& \frac{m_h^2}{4v^2} ( |H_{\rm SM}|^2 - v^2 )^2
    \nonumber \\ &&
    + \frac{1}{8v^2} (2m_W^2 + m_Z^2 + 2m_t^2) T^2 |H_{\rm SM}|^2.
\end{eqnarray}
Minimizing this potential, we obtain $v(T)$; denoting critical
temperature (which is defined as the temperature where the curvature
of $V_T$ at $H_{\rm SM}=0$ vanishes) as $T_C$, we obtain the
expectation value at $T<T_C$ as
\begin{eqnarray}
  v(T) = v \, \sqrt{ 1 - T^2/T_C^2} \,.
\end{eqnarray}
In our calculation, we use $m_{h}$ = 115 GeV and $m_t$ = 175 GeV.
In this case, $T_C = 139$ GeV.

With the shift of the expectation value of Higgs boson at high
temperature, the mixing angle is modified, as shown in Eq.\ 
(\ref{eq:TH_T-dep}).  The effects are from the change of the
left-right mixing mass of sneutrino, and also from the shift of
$m_{\tilde \nu_L}^2(T)$ via the temperature dependence of $D$-term
condensation.  (The second effect was not considered in
Ref.~\cite{Asaka:2005cn}.)  In this paper, we have included both of
these effects and calculated the density parameter of $\tilde{\nu}_R$.

In calculating the temperature-dependent mass of left-handed sneutrino
$m_{\tilde \nu_L}^2(T)$, one may think that it is also necessary to
take account of the thermal mass of the sneutrino.  At the temperature
$T>T_C$, the thermal mass is from the gauge-boson-loop diagrams, and
is estimated as
\begin{eqnarray}
  \label{eq:THMsnul}
    \delta m_{\tilde \nu_L}^2 = 
    \frac{1}{16} (3g_2^2 + g_1^2) T^2.
    \label{T^2mass}
\end{eqnarray}
Importantly, however, we are interested in the case where $v(T)$ is
non-vanishing since the production of $\tilde{\nu}_R$ is effective in
such a case.  Then, the thermal mass of left-handed sneutrino should
be suppressed since gauge bosons acquire masses from the expectation
value of the Higgs boson.  Detailed calculation of the thermal mass is
more complicated in this case and we do not go into the detail.
Instead, we perform our analysis with Eq.\ (\ref{T^2mass}) and with
$\delta m_{\tilde \nu_L}^2=0$ (i.e., with neglecting the thermal mass
of left-handed sneutrino).  As we will see, production of right-handed
sneutrino becomes effective at relatively low temperature and
$\Omega_{\tilde{\nu}_R}^{\rm CE}$ does not change much between two
procedures.

\begin{figure}[t]
    \begin{center}
        \includegraphics[scale=1.3]{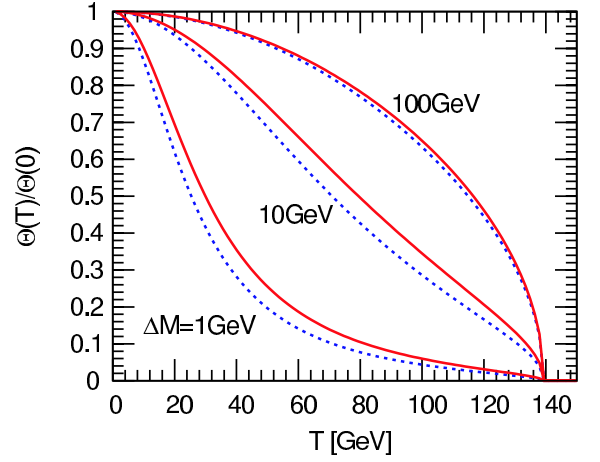}
    \end{center}
    \caption{
    The left-right mixing angle of sneutrinos at finite temperatures
    $\Theta (T)$ normalized bu $\Theta(0)$.  We take here $m_{\tilde
    \nu_R} = 100$ GeV, and the mass difference $\Delta M= m_{\tilde
    \nu_L} - m_{\tilde \nu_R}$ is 1, 10 and 100 GeV from left to
    right. The results obtained by using $\delta m_{\tilde \nu_L}^2=0$
    and $\delta m_{\tilde \nu_L}^2$ given in Eq.\ 
    (\ref{eq:THMsnul}) are shown by the solid and dashed lines,
    respectively.  }
    \label{fig:LRMixing_T}
\end{figure}

In Fig.~\ref{fig:LRMixing_T} we show the left-right mixing angle at
the finite temperature (relative to the zero-temperature value).  We
can see that the mixing angle is drastically suppressed when the
temperature becomes close to $T_C$.

%%%%%%%%%%%%%%%%%%%%%%%%%%%%%%%%%%%%%%%%%%%%%%%%%%%%%%%%%%%%%%%%%%%%%
%%%%%%%%%%%%%%%%%%%%%%%%%%%%%%%%%%%%%%%%%%%%%%%%%%%%%%%%%%%%%%%%%%%%%

\begin{figure}[t]
 \begin{center}
     \includegraphics[scale=1.3]{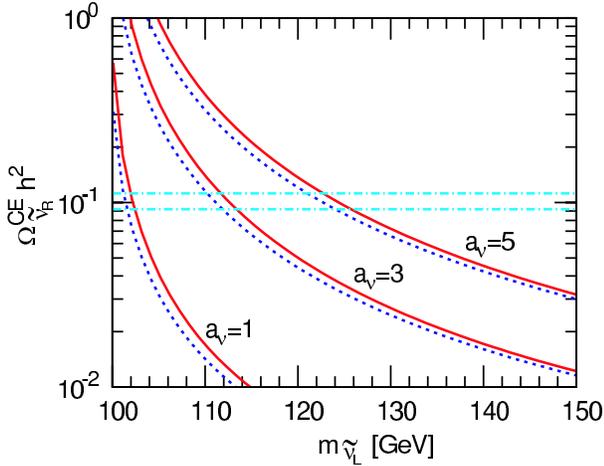}
 \end{center}
 \caption{
   The relic density $\Omega_{\tilde \nu_R}^{\mbox{\tiny CE}} h^2$ of
   the LSP $\tilde \nu_R$ in terms of the left-handed sneutrino mass
   $m_{\tilde \nu_L}$.  The results obtained by using $\delta m_{\tilde
     \nu_L}^2=0$ and $\delta m_{\tilde \nu_L}^2$ given in Eq.\ 
   (\ref{eq:THMsnul}) are shown by the solid and dashed lines,
   respectively.  We take $a_\nu$ = 1, 3, 5 from left to right.  Here
   $m_{\tilde \nu_R}=100$ GeV, $m_{\tilde W}=300$ GeV and $\mu_H = 150$
   GeV.  The horizontal dot-dashed lines correspond to
 the dark-matter density (\ref{eq:omega_dm}).}
 \label{fig:OMcomp}
\end{figure}

\begin{figure}[t]
    \begin{center}
        \includegraphics[scale=1.3]{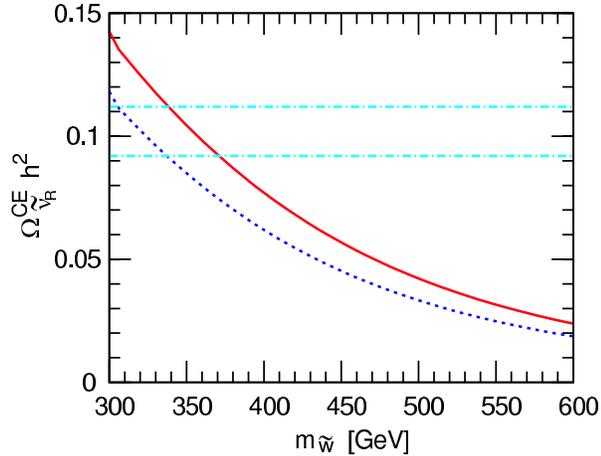}
    \end{center}
    \caption{ 
      The relic density $\Omega_{\tilde \nu_R}^{\mbox{\tiny CE}} h^2$
      of the LSP $\tilde \nu_R$ in terms of the wino mass.  The
      results obtained by using $\delta m_{\tilde \nu_L}^2=0$ and
      $\delta m_{\tilde \nu_L}^2$ given in Eq.\ (\ref{eq:THMsnul}) are
      shown by the solid and dashed lines, respectively.  We take here
      $a_\nu = 3$, $m_{\tilde \nu_R}=100$ GeV, $m_{\tilde \nu_L} =
      110$ GeV and $\mu_H = 150$ GeV.  The horizontal dot-dashed lines
      correspond to the dark-matter density (\ref{eq:omega_dm}).  }
    \label{fig:OMCE_mwino}
\end{figure}

Now, we estimate the relic density of $\tilde \nu_R$ coming from
the decays of superparticle in the chemical equilibrium. We calculate
$\Omega_{\tilde \nu_R}^{\mbox{\tiny CE}}$ by solving Eq.\ 
(\ref{BoltzmannEq}).  In Fig.~\ref{fig:OMcomp}, we show
$\Omega_{\tilde \nu_R}^{\mbox{\tiny CE}} h^2$ as a function of
$m_{\tilde \nu_L}$.  As we have mentioned, $\Omega_{\tilde
\nu_R}^{\mbox{\tiny CE}}$ becomes larger as $m_{\tilde \nu_L}$ gets
close to $m_{\tilde \nu_R}$.\footnote
%%%%
{ Compared with the previous analysis in
Ref.~\cite{Asaka:2005cn}, $\Omega_{\tilde \nu_R}^{\mbox{\tiny CE}}
h^2$ is reduced since the left-right mixing angle of sneutrinos
becomes smaller due to the additional thermal effects.  }
%%%% 
In particular, $\Omega_{\tilde \nu_R}^{\mbox{\tiny CE}}$ becomes
consistent with the density parameter of dark matter with the mass
degeneracy of $\sim 10\ \%$ for $|a_\nu| < 3$.  We have also checked
that $\Omega_{\tilde \nu_R}^{\mbox{\tiny CE}}$ decreases as $m_{\tilde
W}$ increases.  This is due to the fact that the sneutrino production
at higher temperature becomes ineffective since the left-right mixing
angle of sneutrinos is suppressed at high temperature. (See
Fig.~\ref{fig:OMCE_mwino}.)

Eq.\ (\ref{Omega(CE):h}) suggests another possibility of enhancing
$\Omega_{\tilde \nu_R}^{\mbox{\tiny CE}}$.  By assuming degenerate
neutrino mass, Yukawa coupling constants of neutrinos become larger
and hence more $\tilde{\nu}_R$ can be produced.  With Eq.\ 
(\ref{Omega(CE):h}), for example, observed dark-matter density given
in Eq.\ (\ref{eq:omega_dm}) is explained when neutrino masses are
larger than $0.39$ eV with degenerate neutrino masses.  Notice that
this value of the neutrino mass is consistent with the upper bound on
the neutrino mass derived from the WMAP three-year data.

%%%%%%%%%%%%%%%%%%%%%%%%%%%%%%%%%%%%%%%%%%%%%%%%%%%%%%%%%%%%%%%%%%%%%
%%%%%%%%%%%%%%%%%%%%%%%%%%%%%%%%%%%%%%%%%%%%%%%%%%%%%%%%%%%%%%%%%%%%%

%%%%%%%%%%%%%%%%%%%%%%%%%%%%%%%%%%%%%%%%%%%%%%%%%%%%%%%%%%%%%%%%%%%%%
%%%%%%%%%%%%%%%%%%%%%%%%%%%%%%%%%%%%%%%%%%%%%%%%%%%%%%%%%%%%%%%%%%%%%

Up to now, we have discussed the two sources of the $\tilde \nu_R$
relic density, {\it i.e.}, $\Omega_{\tilde \nu_R}^{\mbox{\tiny FO}}$
and $\Omega_{\tilde \nu_R}^{\mbox{\tiny CE}}$.  If both of them are
sizable, the total present relic abundance is given by the sum of
these two contributions:
\begin{eqnarray}
  \Omega_{\tilde \nu_R} 
  =
  \Omega_{\tilde \nu_R}^{\mbox{\tiny FO}}
  +
  \Omega_{\tilde \nu_R}^{\mbox{\tiny CE}}  \,.
\end{eqnarray}

Before closing this subsection, we comment on the case where the
right-handed neutrinos have Majorana mass.  If the Majorana mass is
smaller than the supersymmetry breaking masses of superparticles,
$\tilde{\nu}_R$ may still be the LSP.  In this case, however, the
universe is likely to be overclosed with $\tilde{\nu}_R$-LSP.  This is
because $y_\nu$ becomes larger than the value given in Eq.\ 
(\ref{eq:ynuatm}) with the Majorana mass.  In this case,
$\Omega_{\tilde \nu_R}^{\mbox{\tiny CE}}$ becomes much larger than the
value we obtained for the purely Dirac case unless the reheating
temperature is so low that superparticles can never be thermalized.
(See also \cite{Gopalakrishna:2006kr}.)

%%%%%%%%%%%%%%%%%%%%%%%%%%%%%%%%%%%%%%%%%%%%%%%%%%%%%%%%%%%%%%%%%%%%%%%%
%%%%%%%%%%%%%%%%%%%%%%%%%%%%%%%%%%%%%%%%%%%%%%%%%%%%%%%%%%%%%%%%%%%%%%%%
\subsection{Other possibilities}
\label{sec:others}
%%%%%%%%%%%%%%%%%%%%%%%%%%%%%%%%%%%%%%%%%%%%%%%%%%%%%%%%%%%%%%%%%%%%%%%%
%%%%%%%%%%%%%%%%%%%%%%%%%%%%%%%%%%%%%%%%%%%%%%%%%%%%%%%%%%%%%%%%%%%%%%%%

So far, we have discussed the production of $\tilde \nu_R$ by the
decays of MSSM superparticles.  In the rest of this section, we
briefly comment on other possibilities.

First, $\tilde \nu_R$ is potentially produced by the decays of
non-MSSM superparticles.  One interesting candidate is gravitino.
Since gravitino has only the gravitationally suppressed interactions,
its lifetime is very long and gravitino of mass $\sim 100$ GeV decays
after BBN epoch.  When the primordial abundance of gravitino takes
relevant value, its decay produces the correct amount of $\tilde
\nu_R$ and the $\tilde{\nu}_R$-CDM can be realized.  In usual cases,
decay of the gravitino after BBN is severely constrained by the BBN
constraint \cite{BBN_KKM,Kohri:2005wn}.  However, if the gravitino is
the NLSP and also if it decays only into $\tilde \nu_R+\overline \nu$
(and $\tilde\nu_R^\ast+\nu$), the BBN constraints are significantly
relaxed.\footnote
{Similarly, when gravitino is the LSP and $\tilde \nu_R$ is the NLSP,
the gravitino dark matter may be possible without conflicting 
with the BBN observations.}
This is because the branching ratios of gravitino decaying into
hadrons and charged particles are very small in this case.  We should
note that, even in this case, the decay of the MSSM-LSP into gravitino
is also constrained from BBN.  To avoid it, the MSSM-LSP should
dominantly decay into $\tilde{\nu}_R$ (and something else) with
relatively small hadronic branching ratio or relatively short
lifetime.

In addition, in this paper, we have assumed that the initial abundance
of $\tilde \nu_R$ after the reheating of inflation is negligible.
Note, however, that we cannot exclude the possibility that $\tilde
\nu_R$ is produced at the reheating (or also at the preheating) epoch
by the decay of inflaton.  This is possible when $\tilde \nu_R$
couples to the inflaton although it is highly dependent on the details
of the inflation model.

There is yet another possibility.  $\tilde\nu_R$ might be produced
as a coherent oscillation.  This is because $\tilde \nu_R$ has an
almost flat potential and its initial amplitude after the inflation
might be displaced from the origin.  In this case, if the initial
amplitude is of the order of $10^9$ GeV, such coherent mode of $\tilde
\nu_R$ might be the dark-matter assuming that the reheating of the
inflation completes before the start of the $\tilde \nu_R$
oscillation.

%%%%%%%%%%%%%%%%%%%%%%%%%%%%%%%%%%%%%%%%%%%%%%%%%%%%%%%%%%%%%%%%%%%%
%%%%%%%%%%%%%%%%%%%%%%%%%%%%%%%%%%%%%%%%%%%%%%%%%%%%%%%%%%%%%%%%%%%%
%%%%%%%%%%%%%%%%%%%%%%%%%%%%%%%%%%%%%%%%%%%%%%%%%%%%%%%%%%%%%%%%%%%%
\section{Conclusions}
\label{sec:conclusions}
%%%%%%%%%%%%%%%%%%%%%%%%%%%%%%%%%%%%%%%%%%%%%%%%%%%%%%%%%%%%%%%%%%%%
%%%%%%%%%%%%%%%%%%%%%%%%%%%%%%%%%%%%%%%%%%%%%%%%%%%%%%%%%%%%%%%%%%%%
%%%%%%%%%%%%%%%%%%%%%%%%%%%%%%%%%%%%%%%%%%%%%%%%%%%%%%%%%%%%%%%%%%%%

We have considered the scenario where right-handed sneutrino becomes
the CDM of the universe in the framework that neutrino masses are
purely Dirac-type.  We have investigated the production of $\tilde
\nu_R$ by decays of superparticles in the early universe.  Especially,
the decay of the MSSM-LSP after its freeze-out time have been studied
in detail by using the minimal supergravity model of supersymmetry
breaking.  The NLSP is considered as the MSSM-LSP, $\tilde \chi_1^0$
or $\tilde \tau_1$.

We have found that there is a wide parameter range in which $\tilde
\nu_R$ is the LSP, and also that the relic density of the LSP $\tilde
\nu_R$ coming from the late MSSM-LSP decay can be consistent with the
dark-matter density in such a region.  Assuming the minimal
supergravity model, the mass of the $\tilde \nu_R$ is bounded not only
from below but also from above.  The lower bound comes from the
phenomenological constraints ({\it e.g.}, the Higgs boson mass and the
rate of $b \to s \gamma$) or the requirement of $\tilde \nu_R$ being
the LSP.  On the other hand, the upper bound comes from the
dark-matter density.  In addition, in our framework, the MSSM-LSP is
the MSSM-LSP is the Bino-like lightest neutralino (almost right-handed
lighter stau) when $m_{\tilde \nu_R}$ is small (large).  It should be
noted that these features are insensitive to the neutrino mass
hierarchy.

We have also shown that the decay of the MSSM-LSP receives constraints
from BBN, which leads to the upper bound on the mass of the MSSM-LSP.
Indeed, it forbids the MSSM-LSP with mass $\gg 1$ TeV in any case.  It
has been found that, when $\tan \beta = 10$ and $\mu_H >0$, the 
$\tilde \chi_1^0$ MSSM-LSP can be consistent with the BBN even with
the hierarchical neutrino masses, if the mass of $\tilde \chi_1^0$ is
small enough.  On the other hand, the $\tilde \tau_1$ MSSM-LSP
is severely constrained from BBN argument.  We found that its decay
can be consistent with the BBN only when the neutrino masses are
degenerate and $m_\nu$ is large.  However, the BBN bound strongly
depends on the lifetime and becomes irrelevant when the lifetime is
shorter than about 0.1 sec.  If we go beyond the assumptions used in
our analysis, such a short lifetime can be achieved.  In such a case,
the decay of the $\tilde \tau_1$ MSSM-LSP is cosmologically harmless.

When the CDM is indeed $\tilde \nu_R$, it is not easy to test directly
its existence because of very small neutrino Yukawa coupling
constants.  However, the $\tilde{\nu}_R$-CDM scenario has features
quite its own, and we will have indirect hints from the future
experiments and observations.  Since neutrino masses are purely
Dirac-type, experiments of the neutrinoless double-beta decay should
give null results.  The direct searches for dark matter also should
give null results.  If supersymmetry would be found with the
above-mentioned circumstance, $\tilde{\nu}_R$ would be regarded as a
serious candidate of the LSP and CDM.  In addition, the $\tilde \nu_R$
dark matter scenario may be distinguishable from the conventional
scenario where the $\tilde \chi_1^0$ MSSM-LSP is dark matter.  One
interesting feature of the $\tilde{\nu}_R$-CDM is that the MSSM-LSP
can be charged (or even colored), {\it e.g.}, $\tilde \tau_1$ as we
have considered.  Even if the MSSM-LSP is $\tilde \chi_1^0$, as we
have shown, the mass relations between $m_{\tilde \chi_1^0}$ and
$m_{\tilde \tau_1}$ and also between $m_{\tilde \chi_1^+}$ and
$m_{\tilde \tau_1}$ are significantly different in these two
dark-matter scenarios.  These differences will be testable in the
precise measurements of superparticles at future collider experiments.

\begin{acknowledgments}
    The work was partially supported by the grants-in-aid from the
    Ministry of Education, Science, Sports, and Culture of Japan, 
    Nos.\ 16081202, 17340062 and 18740122 (T.A.), and
    No.\ 15540247 (T.M.).
\end{acknowledgments}

%%%%%%%%%%%%%%%%%%%%%%%%%%%%%%%%%%%%%%%%%%%%%%%%%%%%%%%%%%%%%%%%%%%%
%%%%% ** Reference ** %%%%%%%%%%%%%%%%%%%%%%%%%%%%%%%%%%%%%%%%%%%%%%
%%%%%%%%%%%%%%%%%%%%%%%%%%%%%%%%%%%%%%%%%%%%%%%%%%%%%%%%%%%%%%%%%%%%

%%%%%%%%%%%%%%%%%%%%%%%%%%%%%%%%%%%%%%%%%%%%%%%%%%%%%%%%%%%%%%%%%%%%
%%%%% ** Reference ** %%%%%%%%%%%%%%%%%%%%%%%%%%%%%%%%%%%%%%%%%%%%%%
%%%%%%%%%%%%%%%%%%%%%%%%%%%%%%%%%%%%%%%%%%%%%%%%%%%%%%%%%%%%%%%%%%%%
%%%%%%%%%%%%%%%%%%%%%%%%%%%%%%%%%%%%%%%%%%%%%%%%%%%%%%%%%%%%%%%%%%%%
%%%%%%%%%%%%%%%%%%%%%%%%%%%%%%%%%%%%%%%%%%%%%%%%%%%%%%%%%%%%%%%%%%%%
%%%%%%%%%%%%%%%%%%%%%%%%%%%%%%%%%%%%%%%%%%%%%%%%%%%%%%%%%%%%%%%%%%%%
\end{document}